\documentclass[aps,prb,10pt,twocolumn]{revtex4-1}
\usepackage{amsmath}
\usepackage{amsfonts}
\usepackage{amssymb}
\usepackage{graphicx}
\usepackage{color}  
\usepackage{bbm}
\usepackage{bm}
\usepackage{hyperref}
\hypersetup{
 colorlinks=true,
 linkcolor=blue,
 filecolor=magenta,  
 urlcolor=cyan}
 
\begin{document}

\title{Induced Spin-texture at 3$d$ Transition Metal/Topological Insulator Interfaces}
\author{Slimane Laref$^{1}$}
\email{slimane.laref@kaust.edu.sa}
\author{Sumit Ghosh$^{1}$}
\author{Evgeny Y. Tsymbal$^{2}$}
\author{Aurelien Manchon$^{1}$}
\email{aurelien.manchon@kaust.edu.sa}
\affiliation{$^{1}$King Abdullah University of Science and Technology (KAUST), Physical Science and Engineering Division (PSE), Thuwal 23955-6900, Saudi Arabia\\
{$^{2}$Department of Physics and Astronomy, University of Nebraska, Lincoln, Nebraska 68588, USA}}

\begin{abstract}
While some of the most elegant applications of topological insulators, such as quantum anomalous Hall effect, require the preservation of Dirac surface states in the presence of time-reversal symmetry breaking, other phenomena such as spin-charge conversion rather rely on the ability for these surface states to imprint their spin texture on adjacent magnetic layers. In this work, we investigate the spin-momentum locking of the surface states of a wide range of monolayer transition metals (3$d$-TM) deposited on top of Bi$_{2}$Se$_{3}$ topological insulators using first principles calculations. We find an anticorrelation between the magnetic moment of the 3$d$-TM and the magnitude of the spin-momentum locking {\em induced} by the Dirac surface states. While the magnetic moment is large in the first half of the 3$d$ series, following Hund's rule, the spin-momentum locking is maximum in the second half of the series. We explain this trend as arising from a compromise between intra-atomic magnetic exchange and covalent bonding between the 3$d$-TM overlayer and the Dirac surface states. As a result, while Cr and Mn overlayers can be used successfully for the observation of quantum anomalous Hall effect or the realization of axion insulators, Co and Ni are substantially more efficient for spin-charge conversion effects, e.g. spin-orbit torque and charge pumping.
\end{abstract}

\maketitle

{\em Introduction - }Three-dimensional ${\mathcal Z}_2$ topological insulators \cite{Hasan2010,Qi2011} (TIs), such as (Bi,Sb)$_{2}$(Se,Te)$_{3}$, have attracted substantial consideration in the past decade because of the coexistence of insulating bulk states with topologically protected surface states (see, e.g., Ref. \onlinecite{Hsieh2009}). Remarkably, these surface states exhibit strong spin-momentum locking so that breaking time-reversal symmetry, by either doping with 3$d$ transition metal elements (3$d$-TM) or interfacing with thin magnetic films, activates exotic topological phenomena such as quantized anomalous Hall \cite{Yu2010} and magnetoelectric effects \cite{Qi2008}. This perspective has triggered a vast amount of theoretical \cite{Larson2008,Zhang2013} and experimental\cite{Scholz2012,Ye2015} efforts focusing on the precise description of the structural, magnetic and electronic properties of magnetic TIs. These works aimed at realizing carrier-mediated ferromagnetism with perpendicular magnetic anisotropy in order to induce a gap on the surface Dirac cone while preserving its spin-momentum locking. These efforts led to the experimental observation of quantum anomalous Hall effect \cite{Chang2013}, and to the realization of axion insulating states \cite{Grauer2017}. \par

Besides the realization of exotic quantum phases of matter, the strong spin-momentum locking of the Dirac states presents a thrilling opportunity to achieve large spin-charge conversion. As a matter of fact, strong spin-orbit coupling (SOC) combined with interfacial symmetry breaking unlocks a variety of spintronics effects such as spin Hall and inverse spin galvanic effects \cite{Manchon2015}. When interfaced with magnetic materials, these mechanisms can lead to efficient spin-charge conversion. Topological insulators adjacent to magnetic materials are therefore ideal platforms for the realization of these effects \cite{Garate2010}, i.e., charge pumping \cite{Shiomi2014} and spin-orbit torque \cite{Mellnik2014,Manchon2019}. The recent demonstration of room temperature current-driven magnetization switching \cite{Han2017} and the observation of magnetic textures at these interfaces \cite{Yasuda2017b,Zhang2018} open encouraging perspectives for the exploitation of TIs in spintronics.\par

\begin{figure}[h]
\centering
\includegraphics[scale=0.25]{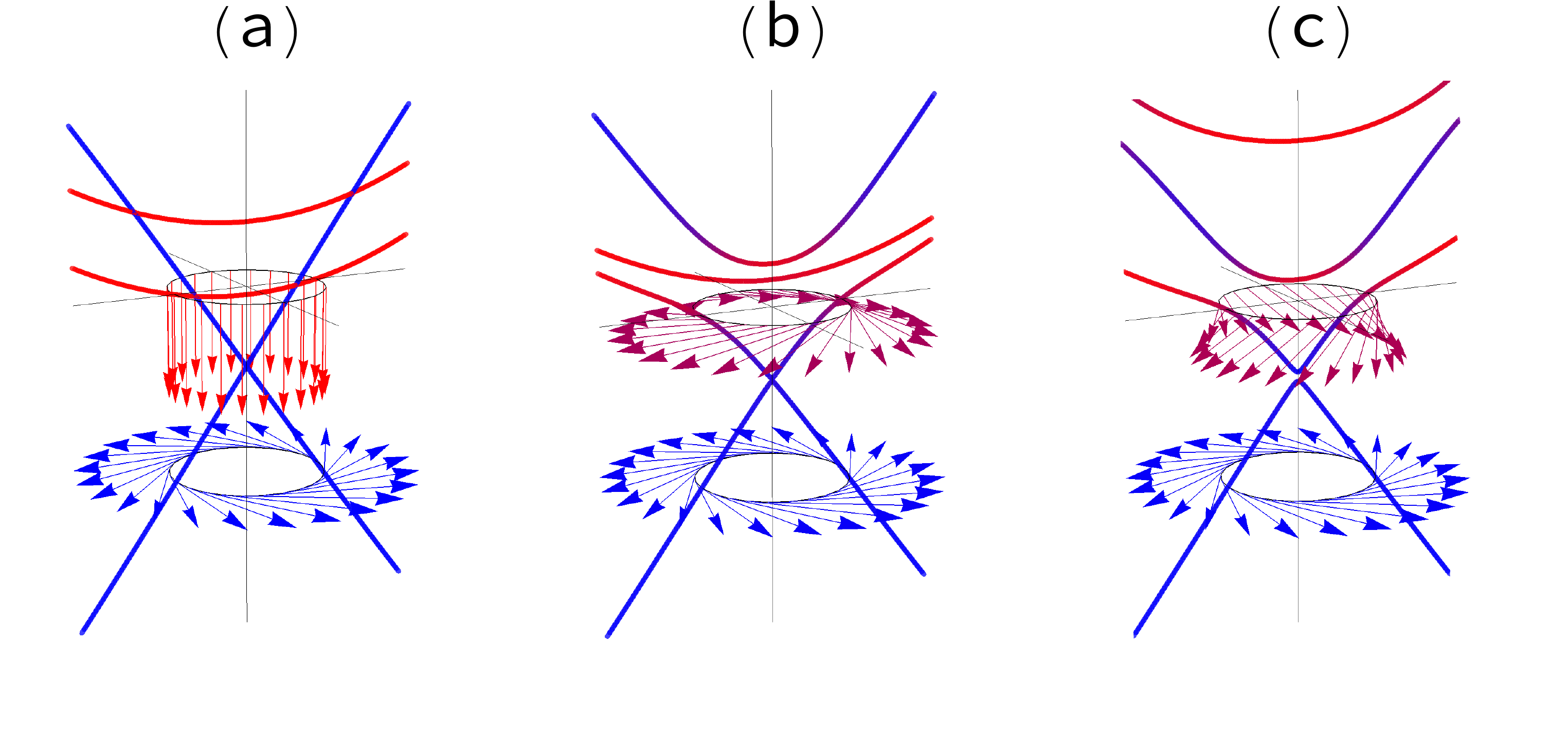}
\caption{(Color online) Band structure of a ferromagnet/TI bilayer as described by Eq. (\ref{eq:1}): (a) uncoupled case ($t$=0), and coupled case ($t\neq$0) with (b) weak and (c) strong exchange $\Delta$. The arrows represent the expectation value of the spin angular momentum operator at different $k$-points in the Brillouin zone, at the interfacial ferromagnet/TI bands. The red and blue colors correspond to the contribution from the ferromagnet and the TI, respectively.}\label{fig1}
\end{figure}

The key physical mechanism underlying efficient spin-charge conversion at 3$d$-TM/TI interfaces is the ability to induce strong spin-momentum locking on the interfacial magnetic states. More specifically, the spin texture needs to be {\em antisymmetric} in momentum $k$, ${\bf S}({\bf k})=-{\bf S}(-{\bf k})$, to enable spin-orbit torque and charge pumping processes \cite{Manchon2015}. Hence, while preserving the Dirac cone and inducing a gap is a crucial ingredient to obtain quantum anomalous Hall effect and axion insulators, it is completely secondary in spin-charge conversion experiments. As a result, while magnetic anisotropy and gap opening have attracted most of the theoretical efforts\cite{Larson2008,Zhang2013}, the nature of spin-momentum locking induced on the 3$d$-TM elements has been only scarcely addressed \cite{Spataru2014, Zhang2016}. Since the 3$d$-TM elements have partially filled 3$d$ states, the bonding character changes considerably across the 3$d$ series, which suggests a strong variation of the induced spin-momentum locking.

In this Letter, we investigate spin-momentum locking at 3$d$-TM/Bi$_2$Se$_3$ interfaces for the complete 3$d$-TM series using density functional theory. We systematically compute the spin texture in momentum space, projected on the 3$d$-TM orbitals, and evaluate its asymmetry in momentum $k$. Remarkably, we find that in the case of magnetic elements in the fcc hollow sites (the most stable configuration), the magnetism is governed by Hund's rule and reaches a maximum for Mn overlayer, while the induced spin-momentum locking exhibits a maximum on Co and Ni overlayers. This anticorrelation between magnetism and induced spin-momentum locking is explained in terms of shell filling and orbital hybridization. This finding sets a guideline for the development of highly efficient spin-charge conversion at the surface of TIs.

{\em Methodology - } Our simulations are based on the pseudopotential plane-wave method with projected augmented wave \cite{Bluchl1994}. Perdew-Burke-Ernzerhof generalized gradient approximation \cite{Perdew1996} is applied through the exchange-correlation functional PBE-D3 dispersion correction \cite{Grimme2010} with Becke-Johnson damping to account for van der Waals corrections, as implemented in the Vienna $ab$ $initio$ Simulation Package (VASP) \cite{Kresse1999}. SOC is self-consistently taken into account in all calculations \cite{Blonski2009}, and we choose a plane-wave cutoff energy of 600 eV. Crystalline Bi$_{2}$Se$_{3}$ has a rhombohedral structure. The 1$\times$1 cell along the (0001) direction is composed of three weakly coupled quintuple layers with a vacuum of at least 20 {\AA}. Since our objective is to assess the amount of spin-momentum locking {\em acquired} by the adjacent 3$d$-TM element by proximity effect, we choose to deposit only one monolayer of 3$d$-TM element on Bi$_{2}$Se$_{3}$ surface. Indeed, while realistic heterostructures involve nanometer-thick 3$d$-TM overlayers \cite{Zhang2016}, proximity effect is expected to affect mostly the magnetic monolayer in contact with the Bi$_{2}$Se$_{3}$ surface. We find that 3$d$-TM monolayer occupies preferentially two different stacking configurations, the magnetic element sitting on fcc and hcp hollow sites \cite{SuppMat} and the former being more stable, consistently with previous studies\cite{Zhang2013}. We also find that the ferromagnetic ordering is favored at the fcc hollow site for Mn, Co and Fe, while Cr is antiferromagnetic and all other elements are nonmagnetic \cite{SuppMat}. In order to allow for systematic comparison across the 3$d$ series, all the calculations below are performed assuming a ferromagnetic ordering and setting the magnetization {\em perpendicular} to the interface in order to facilitate the identification of the momentum-dependence of the spin texture. We emphasize that the fact that the magnetic anisotropy is {\em in-plane} does not invalidate our results. In fact, the magnetic anisotropy is second order in spin-orbit coupling, while the antisymmetric spin-momentum locking is first-order. Therefore, the influence of the magnetic anisotropy on the spin-momentum locking is only of the third order, which is below the accuracy of the numerical simulations [see discussion in Ref. \onlinecite{SuppMat}]. A $\Gamma$-centered 16$\times$16$\times$1 mesh of special $k$ points is adopted for integration over the Brillouin zone. The internal coordinates of all atoms are fully relaxed until the residual forces on each atom are equal or less than 0.01 eV/{\AA}.

{\em Induced spin-momentum locking - } Before discussing the first principles calculations, we turn our attention towards the ideal case of a TI surface interfaced with a ferromagnetic layer (FM). The bilayer is modeled by a 4$\times$4 Hamiltonian,
\begin{equation}\label{eq:1}
{\cal H}={\cal H}_{\rm TI}\otimes(1+\hat{\tau}_z)/2+{\cal H}_{\rm FM}\otimes(1-\hat{\tau}_z)/2+t{\mathbbm 1}_2\otimes\hat{\tau}_x,
\end{equation}
where $\hat{\bm\tau}$ refers to the $\{{\rm TI},{\rm FM}\}$ bilayer subspace, ${\cal H}_{\rm TI}=v\hat{\bm\sigma}\cdot(\hat{\bf p}\times{\bf z})$ is the Dirac Hamiltonian describing the TI surface, and ${\cal H}_{\rm FM}=\hat{\bf p}^2/(2m)+\Delta\hat{\sigma}_z$ represents the FM layer with $s$-$d$ exchange $\Delta$. $\hat{\bm\sigma}$ is the vector of Pauli spin matrices and $t$ is the spin-independent interlayer hopping energy. The spin texture of the Dirac (blue) and magnetic bands (red), ${\bf S}({\bf k})$, is represented in Fig. \ref{fig1}. When the two layers are uncoupled [Fig. \ref{fig1}(a)], the spin texture of the Dirac band adopts the usual Dirac symmetry, ${\bf S}({\bf k})\propto \eta{\bf z}\times{\bf k}$ ($\eta$ is the band index), while the spin texture of the magnetic band is aligned along ${\bf z}$. When the coupling $t$ is turned on, the Dirac and magnetic bands hybridize so that the spin texture of the magnetic bands arises from the competition between (induced) SOC and magnetic exchange. In the case of weak magnetic exchange [Fig. \ref{fig1}(b)], the spin texture of the hybridized magnetic band is close to that of the uncoupled Dirac band, ${\bf S}({\bf k})\propto \eta{\bf z}\times{\bf k}$, while in the case of strong magnetic exchange [Fig. \ref{fig1}(c)], the spin texture is dominated by the exchange, ${\bf S}({\bf k})\propto {\bf z}$. From this toy model, we conclude that the spin-momentum locking {\em induced} on the 3$d$-TM bands, and therefore the spin-charge conversion efficiency, is determined by a compromise between hybridization $t$ and magnetic exchange $\Delta$. The latter is related to magnetic moment $\mu$ through the Stoner criterion, $\Delta = J\mu$, where $J$ is the intra-atomic exchange. Since $J$ varies weakly over the 3$d$ series (from 0.8 eV in V to 1 eV in Ni), $\Delta$ is mostly controlled by the magnetic moment and therefore strongly depends on the 3$d$ orbital filling.\par

\begin{figure}[h]
\centering
\includegraphics[width=8.5cm]{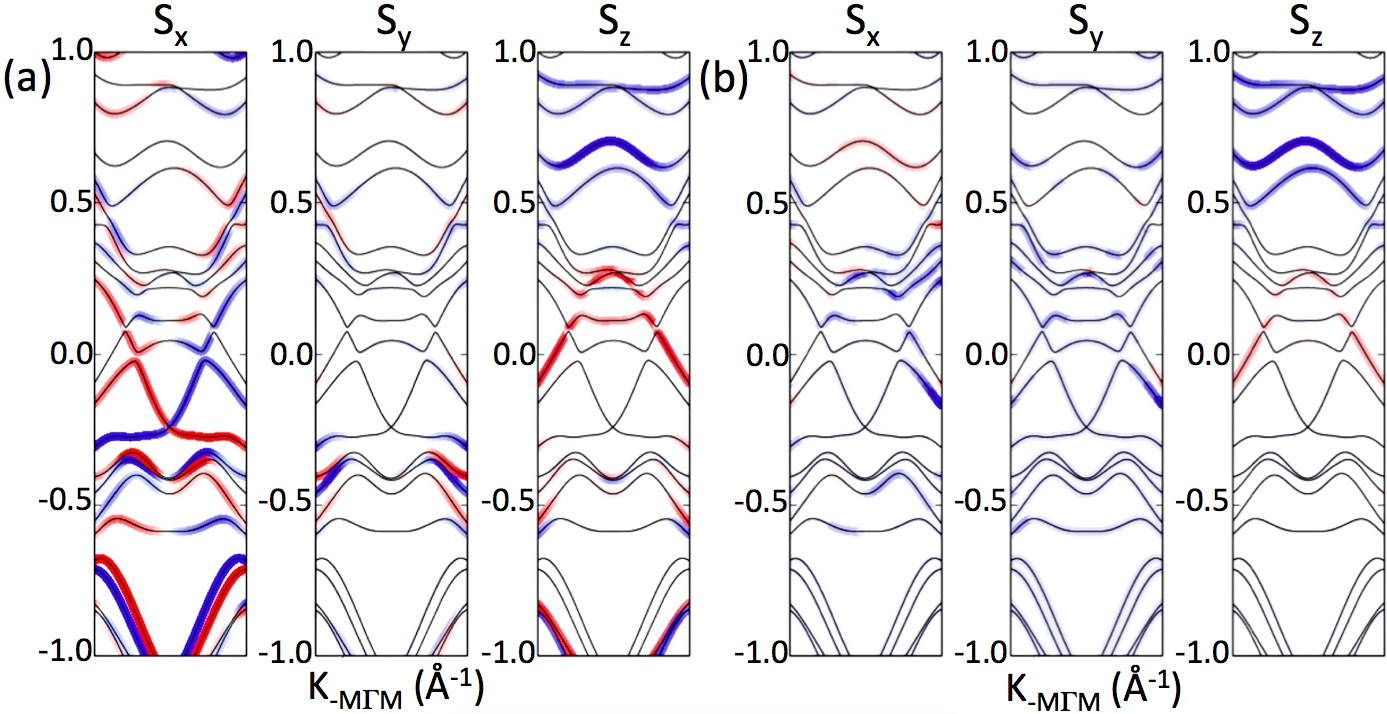}
\caption{(Color online) The relativistic band structure of Mn$_{\rm fcc}$/Bi$_{2}$Se$_{3}$ heterostructure together with the corresponding three components of the spin moment, $S_x$ (left), $S_y$ (center) and $S_z$ (right). Panel (a) displays the total spin moment while panel (b) shows the spin moment projected on Mn-3$d$ orbitals.}\label{fig2}
\end{figure}

We next compute the spin-resolved band structure of 3$d$-TM/Bi$_2$Se$_3$ interfaces for the full 3$d$ series following the method described above. Here, we focus on the spin texture obtained for the fcc hollow site. The results for hcp site are analyzed in Ref. \onlinecite{SuppMat}. We first consider the electronic band structure of Mn$_{\rm fcc}$/Bi$_{2}$Se$_{3}$ slab along $\Gamma{\rm M}$, represented in Fig. \ref{fig2}(a). The Dirac cone, easily identifiable, is shifted down by -0.24 eV, due to the excess of electronic charge brought by the Mn overlayer. We emphasize that this Dirac cone is associated with the {\em bottom} surface and therefore does not hybridize with the 3$d$-TM elements \cite{SuppMat}. The total spin projection along $S_{x}$, $S_{y}$, and $S_{z}$ is represented by the color gradient, from blue (-) to red (+). In this configuration, the out-of-plane spin component is symmetric in momentum, $S_{z}({\bf k})=S_{z}(-{\bf k})$, while the in-plane spin components are antisymmetric in $k$, $S_{x,y}({\bf k})=-S_{x,y}(-{\bf k})$. Hence, the spin-resolved band structure in Fig. \ref{fig2}(a) exhibits a helical spin texture, i.e. ${\bf S}({\bf k})\sim {\bf z}\times{\bf k}$, as expected from the coexistence of large SOC and inversion symmetry breaking. \par

\begin{figure}[h]
\centering
\includegraphics[width=8cm]{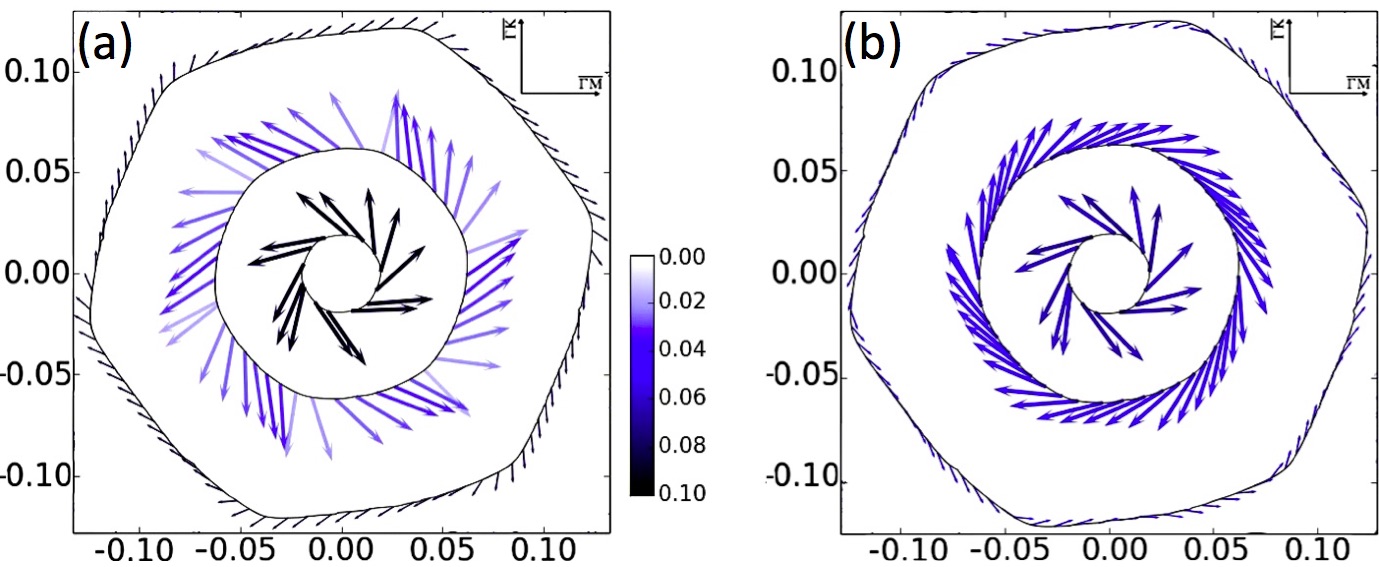}
\caption{(Color online) (a) In-plane total and (b) projected spin texture of Mn$_{\rm fcc}$/Bi$_{2}$Se$_{3}$ heterostructure, represented in momentum space and at Fermi level. The arrows are normalized for better visualization and their color scales with the magnitude of the in-plane projection, $\sim\sqrt{S_x^2+S_y^2}$. The calculation is performed at $E_{\rm F}-0.05$ eV.}\label{fig3}
\end{figure}

It is however difficult at this stage to identify the physical origin of this spin texture, namely to distinguish the contributions of the top and bottom surface and that of 3$d$-TM orbitals. To quantify how much of this spin texture is imprinted on the 3$d$-TM orbitals, Fig. \ref{fig2}(b) presents the spin-resolved band structure {\em projected} on the Mn orbitals. Since Mn $d$-states lie quite far away in energy from the Bi and Se $p$-states (right panel), they hybridize weakly, resulting in a weak induced spin-momentum locking (left and central panels). To complete this analysis, Fig. \ref{fig3} displays the (a) total and (b) projected spin texture close to Fermi energy, $E_{\rm F}$, for the first three bands around the $\Gamma$ point. In this figure, the size of the arrows is normalized for better visualization and the magnitude of the in-plane spin moment, $\sqrt{S_x^2+S_y^2}$, is indicated by the color scale. The first three bands display a circular Fermi surface with slight hexagonal warping and exhibit a spin texture similar to that of the pristine TI surface [Fig. \ref{fig3}(a)]. The spin texture of the inner Fermi contour points in-plane, dominated by Bi$_{2}$Se$_{3}$-$p$ states, while the spin texture of the second band has a much smaller overall magnitude. Based on the analysis of the band structure, the inner Fermi contour is dominated by the {\em bottom} surface. The spin texture of the third band (outer contour) is much smaller and points mostly out-of-plane [see Fig. \ref{fig2}(a), right panel]. When projected on the Mn-3$d$ orbitals [Fig. \ref{fig3}(b)], the spin texture is reduced to a third of its total value while the spin helicity of the second Fermi contour is reversed. This indicates that the Dirac spin-momentum locking of the surface $p$ orbitals has been imprinted on the 3$d$ states of Mn, enabling spin-charge conversion effects to occur.

\begin{figure}[h]
\centering
\includegraphics[width=8.5cm]{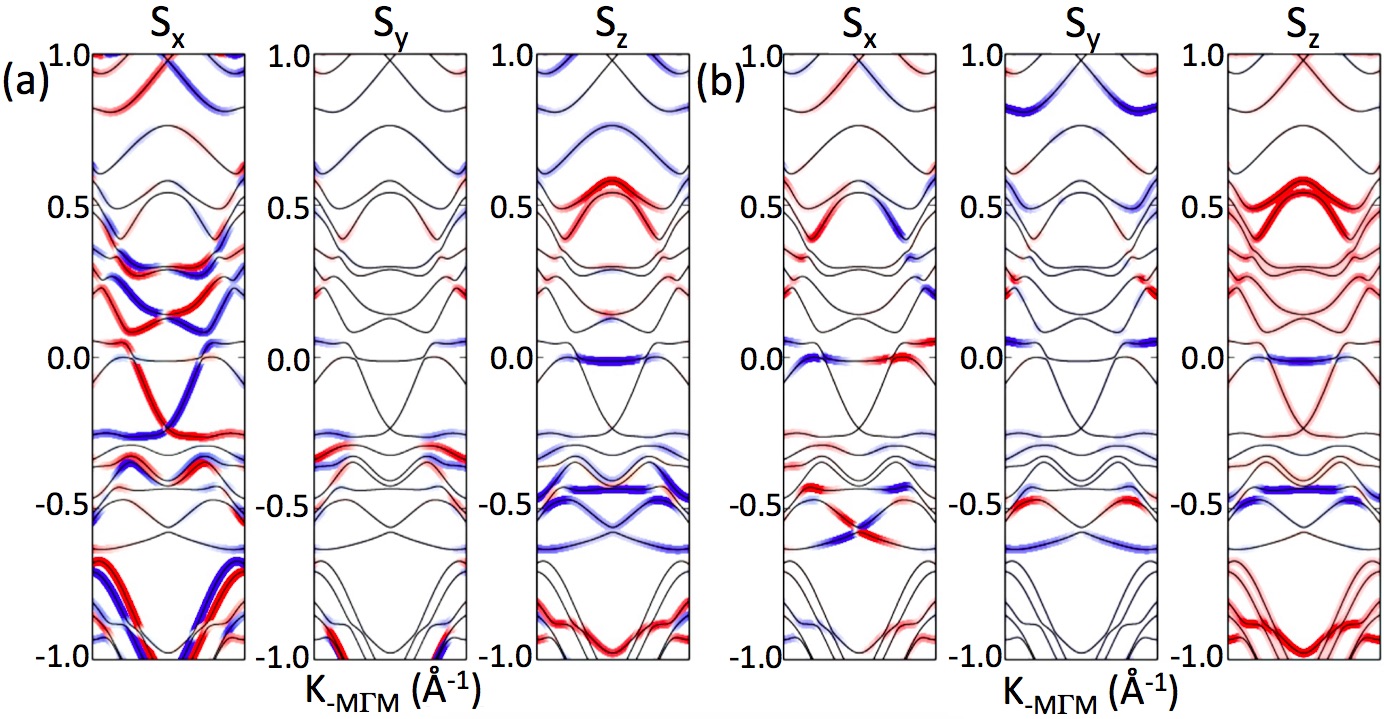}
\caption{(Color online) The relativistic band structure of Co$_{\rm fcc}$/Bi$_{2}$Se$_{3}$ heterostructure together with the corresponding three components of the spin moment, $S_x$ (left), $S_y$ (center) and $S_z$ (right). Panel (a) displays the total spin profile while panel (b) shows the spin profile projected on Co-3$d$ orbitals.}\label{fig4}
\end{figure}

\begin{figure}[h]
\centering
\includegraphics[width=8cm]{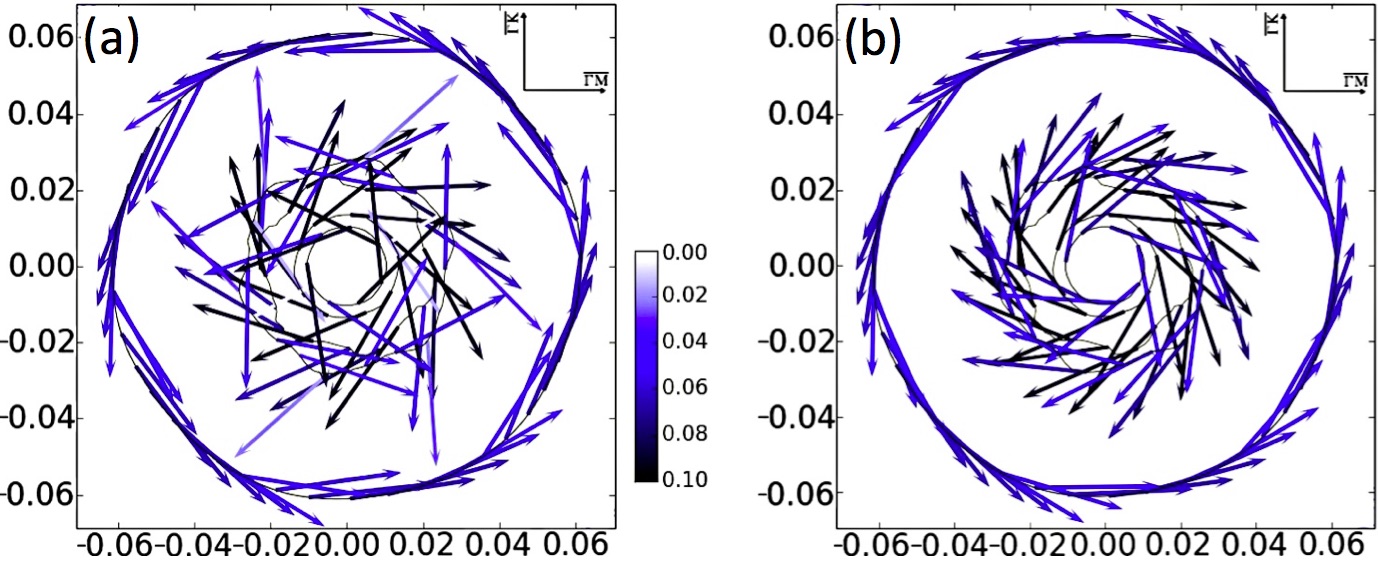}
\caption{(Color online) (a) In-plane total and (b) projected spin texture of Co$_{\rm fcc}$/Bi$_{2}$Se$_{3}$ heterostructure, represented in momentum space and at Fermi level. The representation is the same as in Fig. \ref{fig3}. The calculation is performed at $E_{\rm F}$.}\label{fig5}
\end{figure}

We now move to the case of Co overlayer. The electronic band structure of Co$_{\rm fcc}$/Bi$_2$Se$_3$ slab along $\Gamma{\rm M}$ is represented in Fig. \ref{fig4}(a), together with the total spin projection along $S_{x}$, $S_{y}$, and $S_{z}$ for a magnetization pointing out-of-plane. The Dirac cone associated with the bottom surface is shifted down by -0.23 eV due to charge transfer from the Co layer \cite{Zhang2016}. In this case, the band structure projected on Co 3$d$ orbitals [Fig. \ref{fig4}(b)] reveals that Co states are much closer to Fermi energy and therefore more likely to hybridize with the Bi$_2$Se$_3$ surface states. In fact, the spin texture of Co-3$d$ states is much more pronounced than the spin texture of Mn-3$d$ states. This is evident from Fig. \ref{fig5}(a,b), which reveals the large induced helical spin texture of the Co 3$d$ states. The strong hybridization between Co-3$d$ and Bi$_2$Se$_3$-$p$ states results in several remarkable features. First, the total in-plane spin density exhibits a much more complex texture in the case of Co [Fig. \ref{fig5}(a)] than in the case of Mn  [Fig. \ref{fig3}(a)], due to the strong hybridization between the TI surface states and Co orbitals sitting in the symmetry broken fcc hollow site. Notice that Mn adatom in the fcc hollow site also breaks the symmetry of the Bi$_2$Se$_3$ unit cell, but because Mn hybridizes weakly with the TI surface states, the spin texture of the surface states is weakly perturbed. Second, once projected on Co-3$d$ states, the spin texture adopts an {\em opposite} helicity with respect to the Bi$_2$Se$_3$ pristine surface [see, e.g. the spin helicity in Fig. \ref{fig3}(b)]. This change of spin helicity characterizes proximity-induced Rashba-like spin-momentum locking \cite{Mellnik2014}.

{\em Spin-momentum locking through the 3$d$ series - } Mn and Co overlayers illustrate two opposite paradigms for 3$d$-TM/TI interfaces. In the former, the weak hybridization leaves Bi$_2$Se$_3$-surface states and Mn-3$d$ states mostly unaffected. Hence, when the concentration of Mn element is high enough to induce a gap and the chemical potential is finely tuned, it enables the realization of quantum anomalous Hall effect and axion insulators. In the latter case, the strong $p$-$d$ hybridization modifies the spin texture of the surface states and induces a large helical spin texture on the 3$d$-TM states, unlocking efficient spin-charge conversion. To determine the evolution of the orbital hybridization and the ability of Bi$_2$Se$_3$-surface states to induce spin-momentum locking on the 3$d$-TM overlayer, we systematically compute the induced spin texture at Fermi level of 3$d$-TM$_{\rm fcc}$/Bi$_{2}$Se$_{3}$ heterostructures for all 3$d$ elements, from Sc to Cu.

\begin{figure}[h]
\centering
\includegraphics[width=9cm]{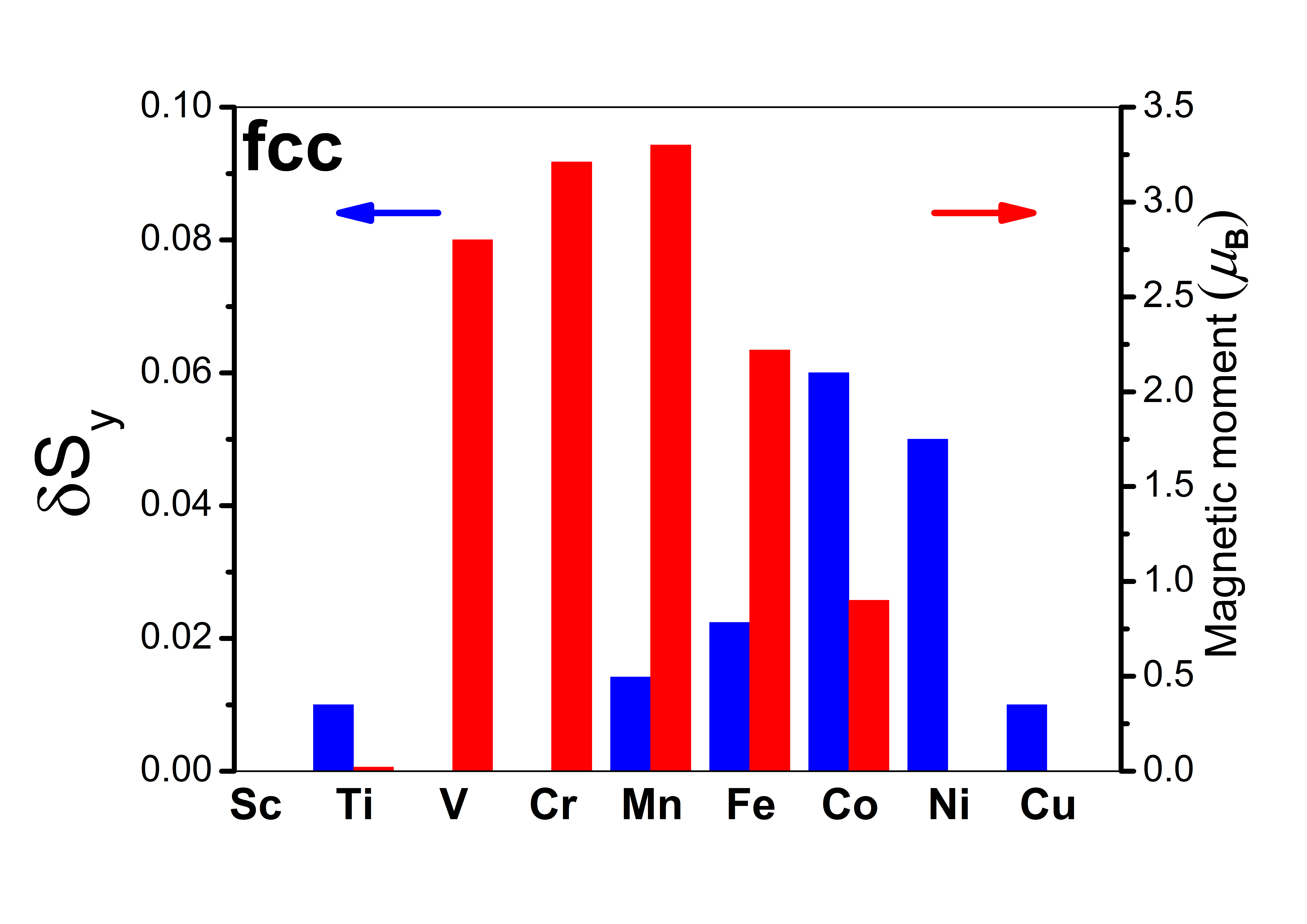}
\caption{(Color online) Induced spin-momentum locking on the 3$d$-TM element (blue histogram - unitless) and total magnetic moment (red histogram - in the units of $\mu_{\rm B}$) of 3$d$-TM/Bi$_2$Se$_3$ slab upon varying the 3$d$-TM overlayer. While the magnetic moment is maximum for Mn overlayer, the induced spin-momentum locking is maximum for Co overlayer.}\label{fig6}
\end{figure}

The 3$d$-TM series is characterized by a progressive filling of the $d$ bands from Sc to Cu. Hence, in thin films the magnetic moment is governed by Hund's rule and reaches a maximum for Mn \cite{Blugel1997,Belabbes2016}. We quantify the induced spin-momentum locking by computing the in-plane spin density projected on the 3$d$-TM element, $\delta S_y=[S_y({\bf k})-S_y(-{\bf k})]/2$, taken on the inner Fermi contour along $\Gamma{\rm M}$. The resulting spin density is reported in Fig. \ref{fig6} for the full 3$d$-TM series (blue), together with the corresponding magnetic moment (red). While the magnetic moment follows Hund's rule and is maximized with Mn overlayer, the induced spin-momentum locking is sizable only for the second part of the 3$d$ series, reaching a maximum for Co overlayer.

{\em Discussion - } This anticorrelation between magnetic moment and induced spin-momentum locking can be understood by inspecting the band structure of the 3$d$-TM/Bi$_2$Se$_3$ series (see Ref. \onlinecite{SuppMat}) and analyzing the hybridization between the 3$d$-TM states and the Se 4$p$ and Bi 6$p$ surface states. A first clue can be deduced from the observation of the magnetic moment (Fig. \ref{fig6}), whose overall trend is similar to the one expected from unsupported 3$d$-TM monolayers \cite{Belabbes2016}: the magnetic moment is maximum around Mn and decreases progressively on both sides. Nonetheless, the trend reported on Fig. \ref{fig6} shows meaningful differences with the monolayer case. While the magnetic moment of V, Cr and, to some extend, Mn overlayers are close to their monolayer values (2.8 vs 3, 3.2 vs 4, and 3.3 vs 4.3 $\mu_{\rm B}$, respectively), the magnetic moment of Fe, Co and Ni overlayers are much smaller (2.2 vs 3.2, 0.9 vs 2.1, and 0 vs 1 $\mu_{\rm B}$, respectively). Although this comparison is only qualitative, it suggests different hybridization schemes for 3$d$-TMs elements with less-than and more-than half-filled $d$-shells.

\begin{figure}[h]
\centering
\includegraphics[width=8.5cm]{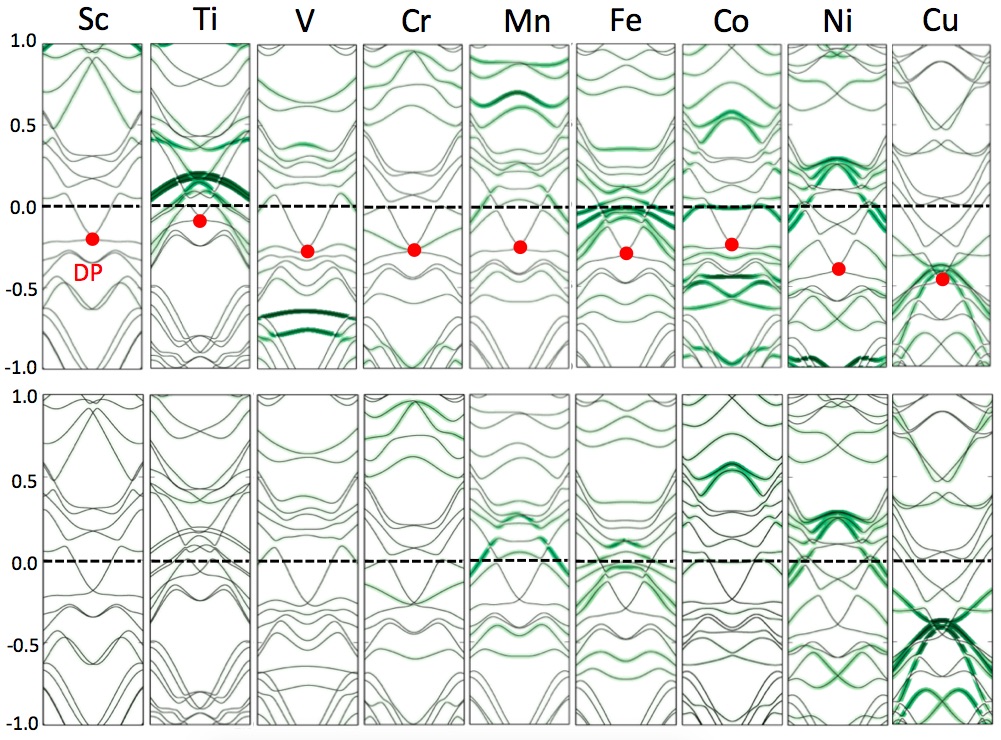}
\caption{(Color online) Band structure of 3$d$-TM/Bi$_2$Se$_3$ for the full 3$d$-TM series, projected on the 3$d$-TM orbitals (top panel) and on the top Se orbitals (bottom panel). The orbital contribution is given by the green shading. The red dot represents the position of the Dirac point (DP) associated with the bottom surface and the dashed horizontal line indicates the position of the Fermi level, $E_{\rm F}$.}\label{fig7}
\end{figure}

This distinct behavior appears more explicitly in Fig. \ref{fig7}, showing the band structures projected on 3$d$-TM orbitals (top panel) and projected on the top Se orbitals (bottom panel). In these figures, the green shading represents the contribution of a specific orbital (3$d$-TM or top Se) to a given band. The darker the shading, the stronger the contribution. The analysis of these band structures reveals that for the initial 3$d$-TM overlayers (from Sc to Cr, with the exception of Ti), the 3$d$ states are located away from the Fermi level, $E_{\rm F}$, thereby reducing the covalent bonding with Se-$p$ states. A similar trend has been reported in the case of (Bi,Sb)$_2$(Te,Se)$_3$ doped by 3$d$-TM elements \cite{Larson2008}. In this case, the weak hybridization prevents substantial induction of spin texture on the 3$d$ states and results in small or vanishing induced spin-momentum locking, as displayed in Fig. \ref{fig6}. In the second half of the 3$d$-TM series (from Mn to Ni), the 3$d$ states are located much closer to Fermi level, thereby enhancing $d$-$p$ hybridization between TM and Se. In other words, covalent bonding starts dominating over exchange energy, and the induced spin texture scales with the hybridization. 

{\em Conclusions - } Employing density functional theory, we demonstrated that the spin texture induced on 3$d$-TM overlayers by Dirac states depends on the interplay between intra-atomic exchange that governs the magnetic moment, and orbital hybridization that scales with the covalent bonding. While the former dominates for 3$d$-TM overlayers with less than half-filled $d$-shell, resulting in a weak induced spin texture, the latter dominates for 3$d$-TM overlayers with more than half-filled $d$-shell, yielding a large induced spin texture. This scenario agrees with the phenomenological picture discussed at the beginning of this Letter. It also suggests that while Cr, Mn and possibly V can be used with success to demonstrate physical effects that require the preservation of Dirac surface states (i.e., quantum anomalous Hall effect and axion insulators), spin-orbit torques and charge-pumping require a completely different situation. Large spin-charge conversion does not necessitate the preservation of Dirac surface states, but rather the induction of large helical spin-momentum locking on the 3$d$ elements, a situation that is optimal in the case Co and Ni. 

\begin{acknowledgments} This work has been supported by the King Abdullah University of Science and Technology (KAUST)
through the Office of Sponsored Research (OSR) [Grant Number OSR-2017-CRG6-3390]. The authors acknowledge computing time on the SHAHEEN supercomputer at KAUST Supercomputing Centre and the team assistance. 
\end{acknowledgments}

\end{document}